\documentclass[12pt,preprint]{aastex}
\usepackage{psfig,rotate}

\shortauthors{Klose et al.}
\shorttitle{Soft Gamma-Ray Repeater 0526--66}

\def\gr{\hbox{ \raisebox{-1.0mm}{$\stackrel{>}{\sim}$} }}

\begin{document}

\title{A near-infrared survey of the N\,49 region around 
the \\ Soft Gamma-Ray Repeater 0526--66\thanks{Based on observations 
       collected at the 
       European Southern Observatory, Paranal, Chile (ESO
       Programme 70.D-0779).}}

\author{
S. Klose\altaffilmark{1},
A. A. Henden\altaffilmark{2},
U. Geppert\altaffilmark{3},
J. Greiner\altaffilmark{4},
H. H. Guetter\altaffilmark{2},
D. H. Hartmann\altaffilmark{5},
C. Kouveliotou\altaffilmark{6},
C. B. Luginbuhl\altaffilmark{2},
B. Stecklum\altaffilmark{1},
F. J. Vrba\altaffilmark{7}
}

\altaffiltext{1}{Th\"uringer Landessternwarte Tautenburg, D--07778 
                 Tautenburg, Germany; e-mail: klose@tls-tautenburg.de}
\altaffiltext{2}{U. S. Naval Observatory/Universities Space Research 
                 Association, Flagstaff Station, Flagstaff, AZ 86001, USA}
\altaffiltext{3}{Astrophysical Institute Potsdam, An der Sternwarte 16,
                 14482 Potsdam, Germany}
\altaffiltext{4}{Max-Planck-Institut 
                 f\"ur extraterrestrische Physik, 85741 Garching, Germany}
\altaffiltext{5}{Department of Physics and Astronomy, Clemson University, 
                 Clemson, SC 29634-0978, USA}
\altaffiltext{6}{NASA/MSFC, NSSTC, SD-50, 320 Sparkman Drive, Huntsville, 
                 AL 35805, USA}
\altaffiltext{7}{U. S. Naval Observatory,
                 Flagstaff Station, Flagstaff, AZ 86001, USA}

\date{Received: 5 April 2004 / Accepted}

%---------------------------------------------------------------

\begin{abstract}
We report the results of a deep near-infrared survey with VLT/ISAAC of the
environment of the supernova remnant N\,49 in the Large Magellanic Cloud,
which contains the soft gamma-ray repeater SGR~$0526-66$. Two of the
four confirmed SGRs are potentially associated with compact stellar clusters.
We thus searched for a similar association of SGR\,0526--66, and imaged a young
stellar cluster at a projected distance of $\sim$30 pc from the
SGR. This constitutes the third cluster-SGR link, and lends support to
scenarios in which SGR progenitors originate in young, dusty clusters. If
confirmed, the cluster-SGR association constrains the age and thus the initial
mass of SGR progenitors.
\end{abstract}


%---------------------------------------------------------------
\section{Introduction}

The neutron star subclass of Soft Gamma-ray Repeaters (SGRs) presently
consists of four confirmed (SGR\,0526$-$66, 1806$-$20, 1900$+$14, 1627$-$41;
see Hurely 2000 and Kouveliotou 2004 for recent reviews) and one candidate 
member (SGR~$1801-23$; Cline et al. 2000). At random intervals, SGRs enter 
active states lasting between days and years, during which they emit
hundreds of predominantly soft ($kT=30$ keV), and short ($0.1-100$ ms
duration) events. During quiescence, SGRs are persistent X-ray sources
($0.1-10$ keV) with luminosities ranging between $\sim10^{33}-10^{36}$ erg
s$^{-1}$. Spin periods, narrowly clustered between $5-8$ s, have been found
in three SGR quiescent X-ray light curves. Estimates of their spin-down rates
indicate that SGR magnetic fields are $10^{14}-10^{15}$ G (Kouveliotou et
al. 1998, 1999; Kulkarni et al. 2003) confirming theoretical predictions 
(Duncan \& Thompson 1992; Thompson \& Duncan 1995) for the existence of such 
high $B$-field objects (magnetars).

Besides SGRs, there is today clear evidence that another class of neutron
stars, Anomalous X-Ray Pulsars (AXPs), possess similar magnetic fields and
SGR-like outbursts (see Mereghetti et al. 2002 for a review; also Kaspi et al.
2004). To date there are roughly ten confirmed magnetars (AXPs and SGRs) and
two to three candidate sources. However, it is still unclear how these two
object  classes are linked, how they produce their unique bursting patterns,
and how  they may be related to the subset of normal radio pulsars that
exhibit magnetic fields comparable or even larger than those of SGRs and AXPs
(see Heyl $\&$  Hernquist 2003 for a recent discussion of these
issues). Studies of possible progenitors for SGRs and AXPs would provide
robust constraints on their ages and their birth rates and thus shed light on
their evolutionary paths. Thus far, associations between magnetars and
Supernova Remnants (SNRs) have been established only for a few AXPs and
potentially in one SGR (Gaensler et al.  2001, and references therein). The
identification of SGRs with fossils of their  births thus remains an open
issue. Three of the four SGRs lie in the Galactic Plane with extinctions of
$10-30$ magnitudes in the optical band. Consequently, infrared is the optimal
band for counterpart searches and for studying their environments. However,
SGR~$0526-66$, resides  in the Large Magellanic Cloud, with very low
extinction ($A_{\rm V} \lesssim 0.1$ mag), allowing for both optical and IR
observations.

Fuchs et al. (1999) studied SGR\,1806$-$20 with ISO and reported the discovery
of a dusty, compact stellar cluster 7$^{\prime\prime}$ away both from the SGR
position and from the Luminous Blue Variable star, LBV\,1806$-$20,  previously
suggested as a potential SGR counterpart (van Kerkwijk et al. 1995).  Corbel
\& Eikenberry (2003) argue that both objects are associated with this
cluster. Recently, Vrba et al. (2000) discovered a similar compact stellar
cluster in the X-ray error box of SGR\,1900+14. This cluster is very close
($\sim0.6$\,pc) to a transient radio source discovered by Frail et al.  (1999)
during the 1998 Giant Flare from the source (Hurley et al. 1999). The infrared
appearance of this cluster is dominated by its most luminous members, two M5
supergiants (Vrba et al. 1996; Guenther, Klose, \& Vrba 2000). These findings
led to the suggestion that both SGRs originated in nearby compact  stellar
clusters (Vrba et al. 2000) and that SGR progenitors may be very massive
stars. To further investigate this hypothesis we focus here on SGR\,0526$-$ 66.

SGR~$0526-66$ was active from the mid 1970s until 1983 and has been in a
quiescent state since then (Aptekar et al. 2001). On 5 March 1979, the source
emitted the most energetic SGR burst ever recorded (Mazets et al. 1979), with
a peak luminosity of over $5\times 10^{44}$ erg s$^{-1}$.  The extreme
intensity and the sharp rise time (0.2\,ms) of this event enabled the first
accurate localization of an SGR source at the edge of the bright SNR N\,49 in
the LMC (Cline et al. 1982). This location was later improved with {\it
Chandra} observations to a 0.6$^{\prime\prime}$ uncertainty (Kulkarni et al.
2003). Kaplan et al. (2001) observed N\,49 with the Hubble Space Telescope but 
could not identify an optical counterpart for the SGR brighter than
$\sim$26.5 mag (F814W filter). 

Here we report on the results of deep near-infrared (NIR) observations of the
N\,49 region using the VLT. While it was not our primary goal to detect the
NIR counterpart of SGR~$0526-66$, we searched for a third cluster-SGR
association,  which would strengthen the case for a physical link between SGRs
and young stellar  clusters of massive stars.

%---------------------------------------------------------------
\section{Observations and data reduction}

We conducted near-infrared imaging of the N\,49 region using VLT/ISAAC in
early 2003 (Table \ref{log}). ISAAC makes use of a 1024$\times$1024 pixel
Rockwell Hg:Cd:Te array and offers a plate scale of 0\farcs147 per pixel.
Because the spectral appearance of N\,49 is characterized by strong Fe
emission lines and a fainter emission component in hydrogen  recombination
lines (Oliva et al. 1989; Dickel et al. 1995), we utilized a combination of
broad- and narrow-band filters for imaging. The narrow-band images (in [Fe II]
at 1.257$\mu$m, [Fe II] at 1.64$\mu$m, and Br$\gamma$ at 2.17$\mu$m) were used
to subtract the nebular contribution from the broad-band images ($J_s, H,
K_s$), allowing the discrimination of continuum sources, i.e. a potentially
hidden stellar cluster, and emission-line knots of the SNR.

All images were analyzed consistently. After standard image processing steps
(flatfielding, stacking), all stars were extracted using DAOPHOT as
implemented  in IRAF\footnote{IRAF is distributed by the National Optical
Astronomical  Observatories, operated  by the Associated  Universities for
Research in Astronomy,  Inc., under contract to the National Science
Foundation}. Psf-fitting was used  with fitting width typically one FWHM
radius. Psf stars were carefully selected   to ensure that no background
contamination was present and that all objects were  stellar. After
extraction, photometry was performed using UKIRT infrared standard  stars FS
6, 14 and 20 observed with VLT/ISAAC.

%---------------------------------------------------------------
\section{Results and Discussion}

\subsection{The NIR view of the N\,49 supernova remnant}

\begin{figure}
\plotone{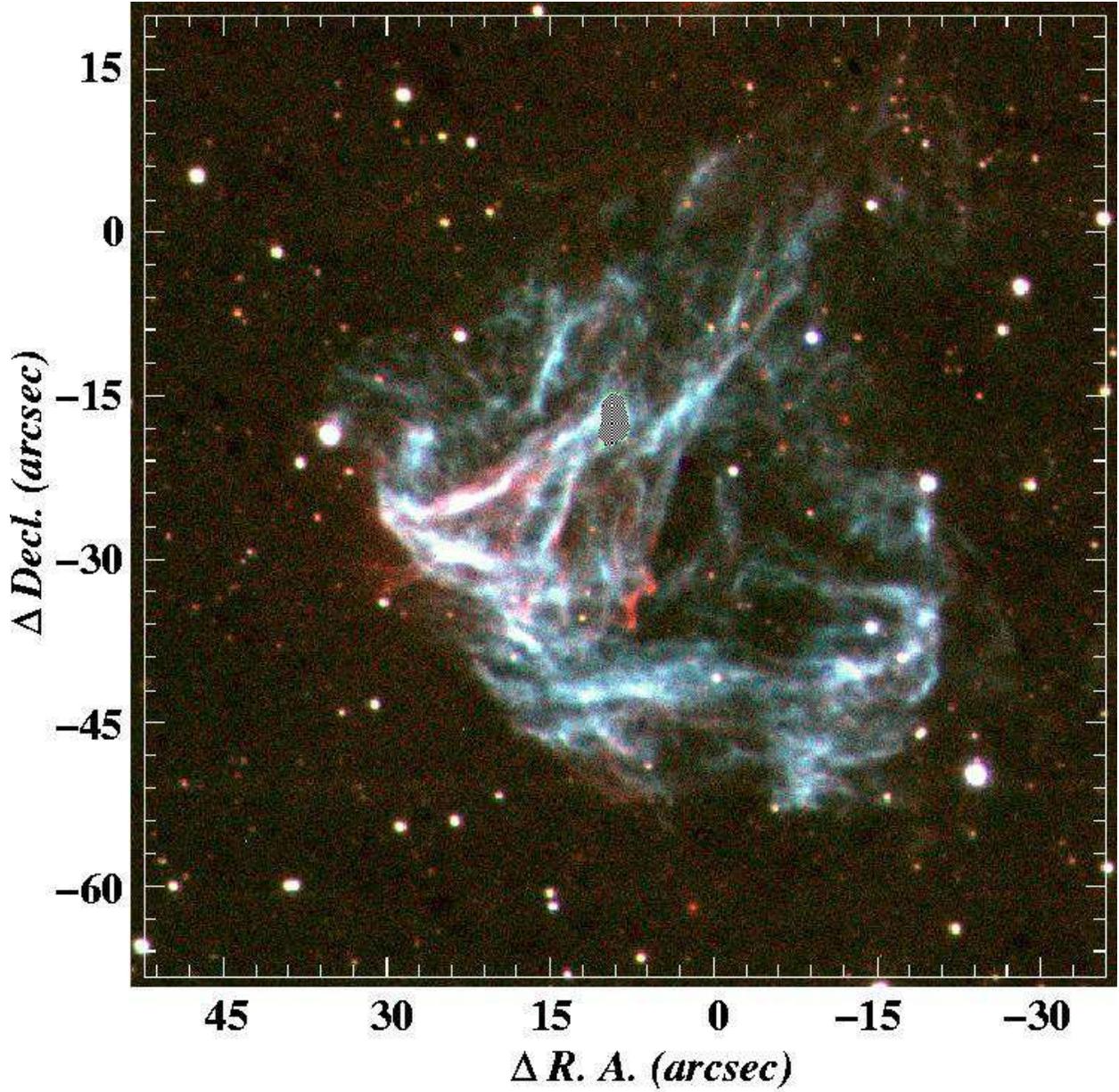}
\caption{$J_sHK_s$ composit of
N\,49, centered at the X-ray position of SGR 0526--66 (Kulkarni et
al. 2003). The figure shows the potential NIR counterpart of IRAS 05259--6607
as a region of excess $K$-band flux (in red color, centered
at $\Delta$ R.A., $\Delta$ Decl. = 15$'', -30''$). Note that the upper right
part of this region is slightly affected by a bad pixel cluster.}
\label{SNR}
\end{figure}

We first searched the NIR data for evidence of a stellar cluster very
close to SGR 0526--66. After removal of the bright line emission from N\,49,
we find no evidence for a cluster in projection against the supernova remnant.
Instead, we find a remarkable excess of $K$-band flux in the south-eastern
part of N\,49 (Fig.~\ref{SNR}). This feature is not seen in our Br$\gamma$
narrow-band image but has a Mid- and Far-Infrared counterpart.

\begin{figure}
\plotone{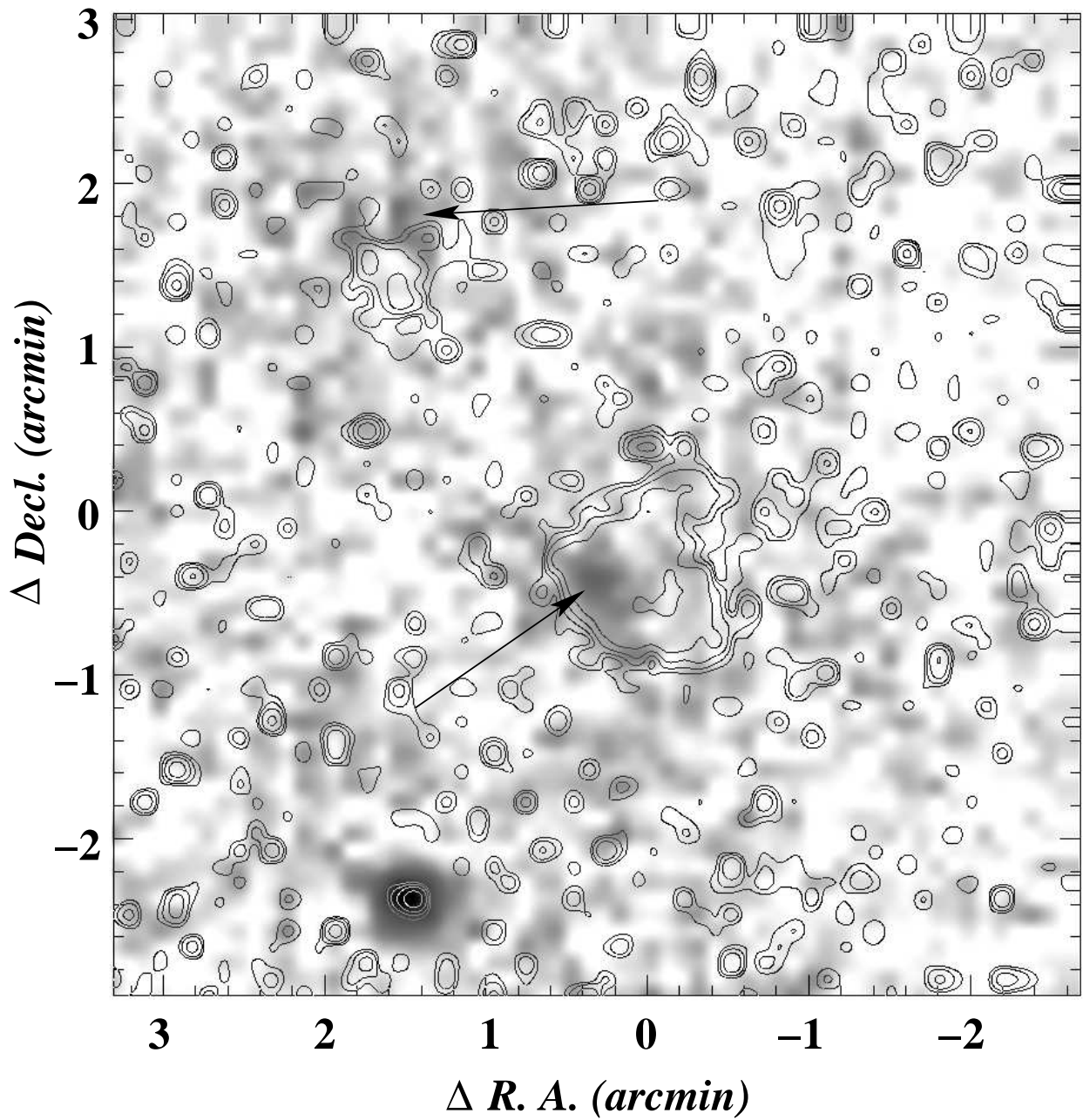}
\caption{MSX view of the N\,49
region at 8.28$\mu$m (see: http://www.ipac.caltech.edu/ipac/msx/) with
contours of the Digitized Sky Survey red plates (DSS2) overplotted, centered
at the X-ray position of SGR 0526--66 (Kulkarni et al. 2003).  While the
south-eastern part of N\,49 appears as a bright source (lower arrow), a
fainter source is visible about  2 arcmin north-east from  the SNR (upper
arrow; see \S~\ref{SL}).  The bright source at the bottom of the image is a
star.}
\label{msx}
\end{figure}

The N\,49 region was observed in the Far-Infrared  with IRAS and in the Mid-IR
by the MSX satellite\footnote{see http://www.ipac.caltech.edu/ipac/msx/}
during  its Galactic Plane survey. The IRAS data were discussed by van
Paradijs et al.  (1996, their figure 1). The brightest source in the field is
recorded in the  IRAS catalog with coordinates R.A., Decl. (B1950) = 05$^{\rm
h}$25$^{\rm m}$59\fs5, --66$^\circ$07$'$03$''$ (Schwering \& Israel 1990),
which dominates at 25$\mu$m and is also bright at 12$\mu$m and 60$\mu$m. The
MSX data of the SNR show a bright and extended source at 8.28$\mu$m with its
center approximately at coordinates R.A., Decl. (J2000) = 5$^{\rm h}$26$^{\rm
m}$03\fs8, --66$^\circ$05$'$05$''$ (Fig.~\ref{msx}), which on our images
coincides with the center of the region where the supernova remnant shows the
excess of $K$-band flux. Since the coordinates of this source/region basically
agree with the coordinates of IRAS 05259--6607 originally published by Graham
et al. (1987), we consider it likely that we have imaged the short-wavelength
counterpart of this IRAS source.
Though it is believed that here the expanding SNR encounters an
interstellar cloud (Banas et al. 1997), based on our data we cannot uniquely
identify the origin of this source and the excess $K$-band flux.

%---------------------------------------------------------------
\subsection{A young stellar cluster close to SGR 0526--66 \label{SL}}

\begin{figure}
\plotone{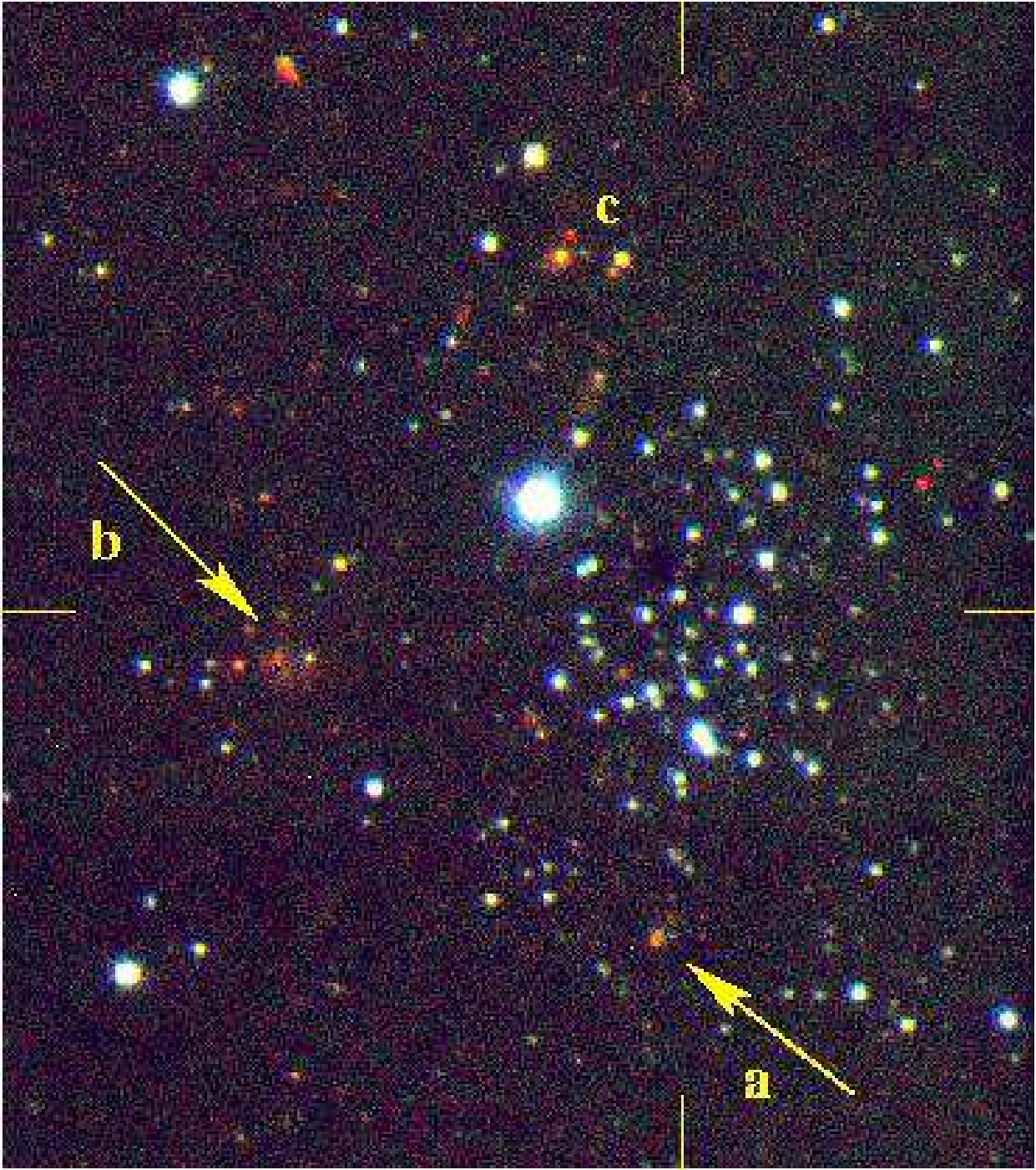}
\caption{$J_sHK_s$ composit of the stellar
cluster SL~463 approximately 130 arcsec north-east from SGR\,0526--66.  The
SGR itself lies outside this image. The coordinate cross marks the suspected
center of the cluster at R.A., Decl. (J2000) = 5$^{\rm h}$26$^{\rm m}$16\fs45,
--66$^\circ$03$'$08\farcs5. Object 'b' is highly reddened and presumably
identical to a bright sub-mm source (Yamaguchi et al. 2001), while another
group of reddened objects ('a') may coincide with an excess flux at 12$\mu$m
seen by IRAS (van Paradijs et al. 1996, their figure 1). A third group of very
red objects ('c')  possibly coincides with an 8.28$\mu$m source
(Fig.~\ref{msx}).  The image size is $\sim75\arcsec\,\times\,65\arcsec$. North
is up, and East is left.}
\label{TheCluster}
\end{figure}

While we have not found a stellar cluster hidden by the bright line emission
of the SNR, a stellar cluster in the vicinity of the SNR does exist. This
cluster is located about 130 arcsec north-east from the quiescent X-ray
counterpart of the SGR (Kulkarni et al. 2003), corresponding to a projected
distance of $\sim$30 pc (Fig.~\ref{TheCluster}). This relatively unexplored
cluster (cataloged  as SL~463 as a member of the OB association LH~53; Hill et
al. 1995; Kontizas et al. 1994) coincides with a bright sub-mm source
(Yamaguchi et al. 2001), indicating that it contains large amounts of gas and
dust. Many objects in this field are highly reddened, indicating on-going
star-formation in this region of the LMC.

\begin{figure}
\plotone{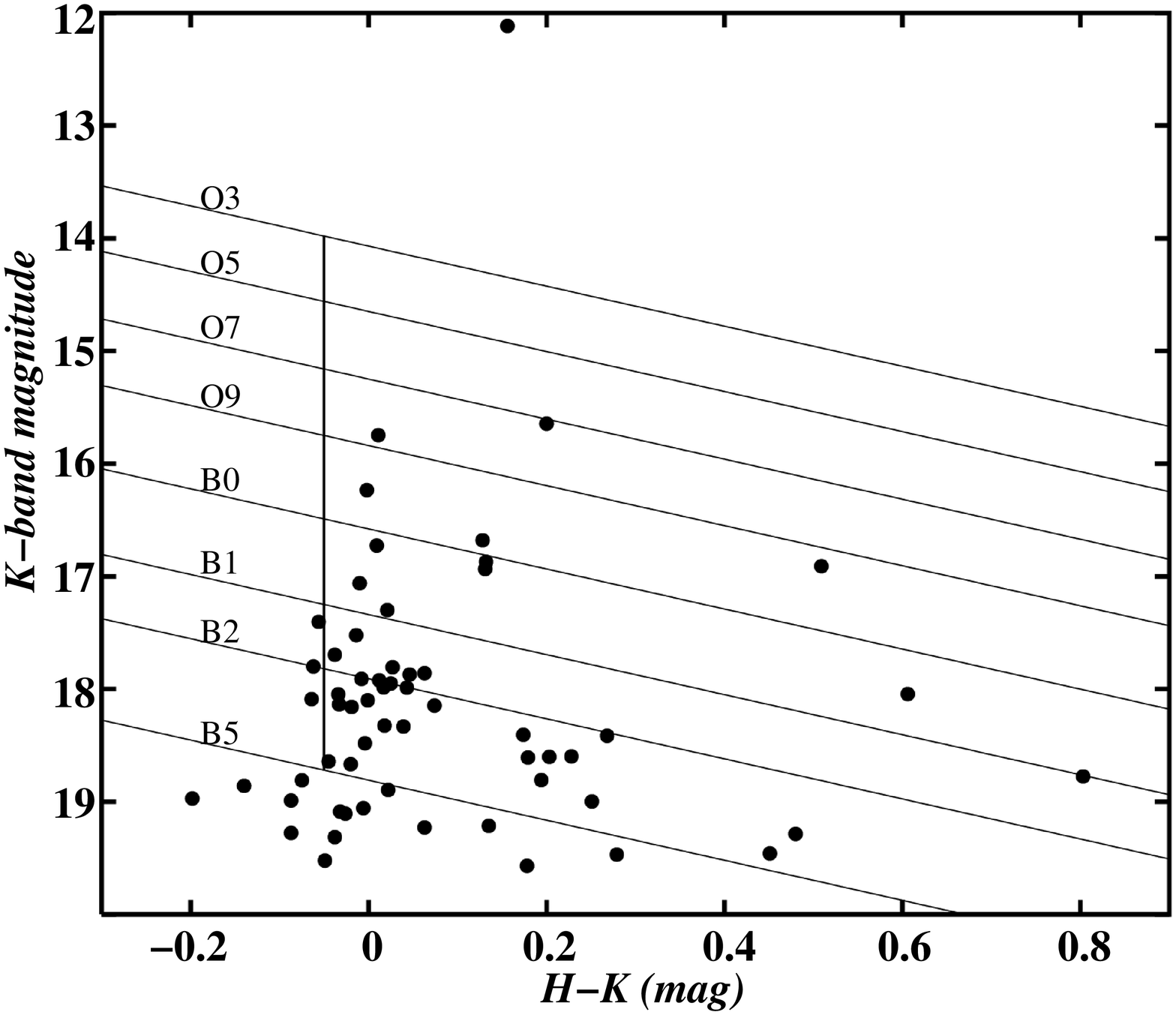}
\caption{Color-magnitude 
diagram for all sources with $JHK$ photometry within $r\le$4.2 pc around the
suspected cluster center. Only stars with accurate photometry (error $<0.1$
mag in $JHK$) are included here. The $K$-band magnitudes of unobscured
main-sequence stars are shown as a vertical line and 
were taken from Hanson, Howarth, \& Conti (1997),  while
setting $H-K=-0.05$ mag (Blum, Conti, \& Damineli 2000), and  assuming a
distance modulus for the LMC of 18.5 mag (Bono et al. 2002). The reddening
tracks for OB main-sequence stars are plotted as straight lines. For a
standard extinction law, 0.2 mag change in $H-K$ color correspond to 3.2 mag
visual extinction  (Rieke \& Lebofsky 1985).}
\label{CMD}
\end{figure}

The $K$ vs. $H-K$ color-magnitude diagram of this cluster indicates the
presence of several dozen B-type main-sequence stars and possibly one
supergiant (Fig.~\ref{CMD}), although we cannot exclude that this is a
Galactic foreground star (the brightest star in Fig~\ref{TheCluster}). Most
members of SL~463 are extinct by less than 1 magnitude, although several stars
might be affected  by somewhat stronger exinction. Within a radius $r=2.1$ pc
of the suspected cluster center (for an assumed distance of 50 kpc), fifty
sources are visible on our ISAAC images. For thirty one of them we have
accurate photometry to a limiting magnitude of 19.5 mag in $JHK$. In the outer
regions of the cluster, between 2.1 and  4.2 pc, an additional ninety stars
are detected. Based on a potential relation between the radius of a young
stellar cluster and its age (Ma\'iz-Apell\'aniz 2001), the age of SL~463 is
between 5 and 20 Myr, in agreement with age estimates of the entire
LH~53 complex (Hill et al. 1995; Yamaguchi et al. 2001).

Compared to the clusters found in close projection to SGR\,1900+14 and
1806--20, the cluster SL~463 is older and larger, but apparently still
enshrouded by dust, and located much farther away from the corresponding SGR.
With respect to the origin of SGR\,0526--66 we briefly consider two
scenarios. \it First, \rm  assuming the cluster was the birthplace of the SGR
progenitor, the SGR must have been ejected from the cluster with a space
velocity of $\sim30\,(\sin\theta)^{-1} n^{-1}$ km s$^{-1}$ in order to travel
a projected distance of 30 pc within $n$ Myr, where $\theta$ is the angle
between the line of sight and the moving direction of the SGR. If the SGR
progenitor and the stellar cluster are coeveal, a cluster age of $\sim$10 Myr
would imply an initial mass of this star of $>20\,$M$_\odot$ (Hill et
al. 1995). Numerical models of the dissolution of young massive clusters have
shown that an ejection of stars with initial masses of 5 to 10\,M$_\odot$ with
velocities of the order of 10 to 40 km s$^{-1}$ indeed occurs (Vine \& Bonnell
2003), if the gas-to-stellar mass ratio in the cluster is relatively high (as
it seems to be the case for SL~463). \it Second, \rm if SGR 0526--66 was born
as a magnetar within this stellar cluster, given its current offset from the
cluster and allowing for an ejection velocity of order
300\,$(\sin\theta)^{-1}$ km s$^{-1}$ its age must be about $10^5$
yrs. This predicts a proper motion of the SGR of $\sim$ 1 milliarcsec
$(\sin\theta)^{-1}$ per year along a direction away from the stellar cluster.

A field strength of about $7\,\times\,10^{14}$ G was inferred from
the $P-\dot{P}$ relation for the quiescent X-ray counterpart of the SGR
(Kulkarni et al. 2003). If the magnetar is coeval with SNR N49 ($\sim$ 5000
years), such a field strength would be consistent with the magnetar age. On
the other hand, for an age of 10$^5$ years required by an SGR-cluster
association, the current $B$-field depends on the assumed evolution model. If
the field evolves through crustal ohmic decay, perhaps accelerated by a Hall 
cascade, irrespective of the initial field strength the remaining field after 
10$^5$ years will be $\sim$ 10$^{13}$ G, which is inconsistent with the value
inferred from the $P$-$\dot{P}$ relation. However, if the field is anchored
in the core and evolves via ambipolar diffusion, then a surface field strength
of $7\,\times\,10^{14}$ G after 10$^5$ years is easily conceivable. Under
certain conditions, magnetars may be able to sustain high field strengths
over such a long period of time (Colpi, Geppert, \& Page 2000). In other
words, the observed field characteristics of the quiescent X-ray counterpart
of the SGR do not exclude the possibility that the birth of the SGR took place 
within the stellar cluster SL~463.

%---------------------------------------------------------------
\section{Conclusions}

We performed a NIR survey towards the N\,49 region in the LMC, and address the
question of the circumstances of the formation of SGR 0526--66. We imaged  the
young and presumably dusty stellar cluster SL~463, located 
only $\sim30$ pc away (projected distance) 
from the SGR. It is possible that SGR 0526--66 was born in this cluster,
but our observations do not allow us to claim with certainty that the SGR or
its progenitor was born within this cluster. However, the fact that similar
clusters have been found at or near the positions of the three best-studied
SGRs  (1900+14, 1806--20, and 0526--66) argues in favor of a cluster/SGR
connection.  If this association is real, we can constrain the age, and thus
the initial mass  of the SGR progenitor. In all three cases the masses turn
out to be \gr20\,M$_\odot$.  This would place SGR progenitors among the most
massive stars with solar metallicity  that can produce neutron star remnants
(Heger et al. 2003).

%---------------------------------------------------------------
\begin{acknowledgements}
We are highly indebted to the ESO staff at Paranal for performing the
observations in service mode. This research made use of data products from
the Midcourse Space Experiment. Processing of the data was funded by the
Ballistic Missile Defense Organization with additional support from NASA
Office of Space Science. This research has also made use of the NASA/IPAC
Infrared Science Archive, which is operated by the Jet Propulsion Laboratory,
California Institute of Technology, under contract with the National
Aeronautics and Space Administration. We thank the referee for a rapid
reply.
\end{acknowledgements}

%---------------------------------------------------------------

%---------------------------------------------------------------
\clearpage

\begin{table}
\caption{Observing log of the N\,49 region with VLT/ISAAC \label{log}}
\vspace{0.3cm}
\begin{tabular}{lll}
\hline
\noalign{\smallskip}
Date (UT) & Filter & Exposure (sec) \\
\noalign{\smallskip}
\hline
\noalign{\smallskip}
 2003 Jan 26 & $J_{\rm s}$    & 14$\times$100 \\
 2003 Jan 26 & $H$            & 8$\times$95  \\
 2003 Feb 17 & $H$            & 6$\times$95  \\
 2003 Feb 17 & $K_{\rm s}$    & 14$\times$75  \\
 2003 Mar 17 & NB 1.26~$\mu$m & 14$\times$160 \\
 2003 Feb 17 & NB 1.64~$\mu$m & 14$\times$160 \\
 2003 Feb 17 & NB 2.17~$\mu$m & 14$\times$160 \\
\noalign{\smallskip}
\hline
\end{tabular}
\end{table}

%---------------------------------------------------------------
\clearpage

%---------------------------------------------------------------
\clearpage

%---------------------------------------------------------------
\clearpage

%---------------------------------------------------------------
\clearpage

\begin{references}

\reference{} Aptekar, R. L., et al. 2001, ApJS, 137, 227 
\reference{} Banas, K. R., et al. 1997, ApJ, 480, 607 
\reference{} Blum, R. D., Conti, P. S., \& Damineli, A. 2000, AJ, 119, 1860 
\reference{} Bono, G., et al. 2002, ApJ, 574, L33
\reference{} Cline, T. L., et al. 1982, ApJ, 255, L 58
\reference{} Cline, T. L., et al. 2000, ApJ, 531, 407
\reference{} Colpi, M., Geppert, U., \& Page, D. 2000, ApJ, 529, L29
\reference{} Corbel, S., \& Eikenberry, S. S. 2003, A\&A, submitted
             (preprint: astro-ph/0311313)
\reference{} Dickel, J., et al. 1995, ApJ, 448, 623                    
\reference{} Duncan, R. C., \& Thompson, C. 1992, ApJ, 392, L9
\reference{} Frail, D. A., et al. 1999, Nature, 398, 127           
\reference{} Fuchs, Y., et al. 1999, A\&A, 350, 891                         
\reference{} Gaensler, B. M., et al. 2001, ApJ, 559, 963
\reference{} Graham, J. R., et al. 1987, ApJ, 319, 126
\reference{} Guenther, E., Klose, S., \& Vrba, F. 2000, 
             in AIP Conf. Proc. 526, 5th Huntsville Symposium on GRBs, ed.
             R. M. Kippen, R. S. Mallozi, \& G. J. Fishman (Melville, New
             York: AIP), 825
\reference{} Hanson, M. M., Howarth, I. D., Conti, P. S. 1997, ApJ, 489, 698 
\reference{} Heger, A. et al. 2003, ApJ, 591, 288
\reference{} Heyl, J. S. \& Hernquist, L. 2003, astro-ph/0312608
\reference{} Hill, R. S., et al. 1995, ApJ, 446, 622
\reference{} Hurley, K., et al. 1999, Nature, 397, 41
\reference{} Hurley, K. 2000, in in AIP Conf. Proc. 510, The Fifth Compton 
             Symposium, ed. M. L. McConnell \& J. M. Ryan, (Melville, New
             York: AIP), 515
\reference{} Kaplan, D. L., et al. 2001, ApJ, 556, 399
\reference{} Kaspi, V. M. 2004, in IAU Symp. 218,
             Young Neutron Stars and Their Environments, 
             ed. F. Camilo \& B. M. Gaensler,
             in press (preprint: astro-ph/0402175)
\reference{} Kontizas, E., et al. 1994, A\&ASS, 107, 77
\reference{} Kouveliotou, C., et al. 1998, Nature, 393, 235
\reference{} Kouveliotou, C., et al. 1999, ApJ, 510, L115
\reference{} Kouveliotou, C. 2004, in ASP Conf. Proc., 
             From X-ray Binaries to Gamma-Ray Bursts,
             ed. E.P.J. van den Heuvel et al., in press
\reference{} Kulkarni, S., et al. 2003, ApJ, 585, 948 
\reference{} Ma\'iz-Apell\'aniz, J. 2001, ApJ, 563, 151
\reference{} Mazets, E. P., et al. 1979, Nature, 282, 587
\reference{} Mereghetti, S., et al. 2002, in MPE Report 278,
             Neutron Stars, Pulsars, \&
             Supernova Remnants, ed. W. Becker et al., 29
\reference{} Oliva, E., et al. 1989, A\&A, 214, 307         
\reference{} Rieke, G. H., \& Lebofsky, M. J. 1985, ApJ, 288, 618 
\reference{} Schwering, P. B. W., \& Israel, F. P. 1990, Atlas and 
             Catalogue of Infrared Sources in the Magellanic Clouds 
             (Dordrecht: Kluwer)
\reference{} Thompson, C. \& Duncan, R. C. 1995, MNRAS 275, 255
\reference{} van Kerkwijk, M. H., et al. 1995, ApJ, 444, L33
\reference{} van Paradijs, J., et al. 1996, A\&A, 299, L41   
\reference{} Vine, S. G., \& Bonnell, I. A. 2003, MNRAS, 342, 314
\reference{} Vrba, F. J., et al. 1996, ApJ, 468, 225   
\reference{} Vrba, F. J., et al. 2000, ApJ, 533, L17      
\reference{} Yamaguchi, R., et al. 2001, ApJ, 553, L185

\end{references}
\end{document}